# Improving the Signal-to-Noise Ratio of Atomic Transition Peaks in LIBS Using Two-dimensional Correlation Analysis

**Linga Murthy Narlagiri, S. Venugopal Rao***

*Advanced Center of Research in High Energy Materials (ACRHEM),
University of Hyderabad, Hyderabad 500046, Telangana, India*
**Corresponding author e-mail:** soma_venu@uohyd.ac.in OR soma_venu@yahoo.ac.in

**Abstract**

In this study, two-dimensional (2D) correlation analysis was utilized for achieving a significant improvement in the signal-to-noise (S/N) ratio of laser-induced breakdown spectroscopy (LIBS) data. Time-resolved LIBS spectra of metallic, bimetallic targets and the normal LIBS spectra of bimetallic targets with varying compositions were used for the detailed analysis. The diagonal of the matrix in the synchronous spectra was used to demonstrate the improvement in the S/N ratio. An improvement in the peak intensities by few orders of magnitude accompanied by suppression in the noise was observed. The correlations between LIBS peaks were also visualized using 2-D plots. Correlation strengths of atomic transitions were visualized in Aluminium (Al), Copper (Cu), and Brass whereas correlation strengths of atomic, atomic and ionic transitions were visualized in Au-Ag bimetallic targets with different compositions (Au30Ag70, Au50Ag50, Au80Ag20). The improved spectra were subsequently used in the principal component analysis for classification studies of four compositions of bimetallic targets (Au20Ag80, Au30Ag70, Au50Ag50, and Au80Ag20). The variance of the first three principal components was found to be improved from the analysis. The accumulated percentage of explained variance of ~95 was achieved with the first three components from improved spectra whereas only ~80 was achieved with the regular LIBS spectra from PCA studies.

**Introduction:**

Laser induced breakdown spectroscopy (LIBS) analysis employs the spectral emissions from the plasma formed when intense laser pulses are focused on to the sample.[1] Typical LIBS spectra consist of ionic, atomic, and molecular transitions.[2] LIBS technique has the ability to identify individual elements in any material. Furthermore, the information in the varying intensities with different composition of elements coupled with machine learning techniques can be used for classification and identification of samples. Because of its versatile nature, being quick and minimal sample preparation requirements, LIBS found fantastic applications in diverse fields such as bacteria classification[3-5], to geological materials[6] in planetary exploration, minerals,[7] archeology,[8,9] explosive detection,[10,11], trace element detection,[12,13] in the study of historic paintings,[14] and in the study of fundamental plasma properties.[15] This technique also found diverse applications because of its suitability to combine with Raman spectroscopy[16,17] to be able to operate easily in double-pulse and/or stand-off mode[18]. The advantage of signal enhancement using nanoparticles[19] and adopting the new trends in machine learning for the LIBS data analysis makes it a powerful tool for different application scenarios. The LIBS spectra of organic materials and few metals contain molecular emissions along with atomic and ionic emissions.[20-22] These molecular emissions can be used to understand the material's properties.[23, 24] Two-dimensional correlation spectroscopy (2DCOS) was initially developed by Noda et al..[25] This technique was primarily used to analyze the data from nuclear magnetic resonance (NMR), near-infrared (NIR), and Raman spectroscopy.[25-28] The resolution of the spectrum was shown to be improved with the 2D spectroscopy.[29] The simultaneous variations or coupling between the two spectral emissions with a small perturbation in the system can be studied. Better interpretation capabilities can be achieved when compared to regular spectra from the shapes and strengths in the 2D correlation plot.[30] Here we have used it to visualize and interpret the correlations in the atomic transitions of Aluminium, Copper, Brass, and a bimetallic target of Au-Ag.

2D correlation technique has been applied to the nanosecond (ns) LIBS data of Aluminium, Copper, Brass, and Au-Ag bimetallic targets. We have demonstrated the 2D correlation spectroscopy application on time-resolved and composition varying LIBS spectra. The analysis can be visualized using two-dimensional contour plots. The strength of the points on diagonal points of the two-dimensional plot represents the self-correlation (or the auto correlation) of the peaks. It provides the information on the time correlation of a particular peak intensity with itself. The strength of the off-diagonal points represents the time correlation of the peak intensity with other peaks. Similarly, the 2D analysis on the spectra with varying concentrations offer information on the auto correlation and cross correlation with composition. Apart from visualization this analysis could also be used in improving the resolution and signal-to-noise ratio (SNR) of the spectrum.

Our detailed analysis confirmed that at least 3-4 orders of magnitude improvement in the signal-to-noise (SNR) of the LIBS data can be achieved using this technique. It is pertinent to note from the collected data/analysis performed that the observed improvements were (a) not limited to few peaks or (b) was not random and it was observed on the complete LIBS spectrum. We have also demonstrated that improved spectra can further be used for improving the classification capabilities (here we used the PCA analysis). This can find impending applications in stand-off LIBS case where the SNR is generally very poor. The perturbation has been achieved by changing the acquisition time and composition of the material. Though this technique has been well established in the improvement/analysis of NMR and IR data we have demonstrated this for the first time using nanosecond LIBS data, to the best of our knowledge. We have also discussed few prospective applications in the conclusions section.

**Experimental Procedure:**

Figure 1 (a) depicts the experimental schematic of the nanosecond LIBS setup. A nanosecond Nd:YAG laser of 7 ns pulse duration at 532 nm with an average pulse energy of 42 mJ was used for the breakdown of the samples. The LIBS signal collected was fed to an ICCD and spectrometer combination. The LIBS spectra of Au-Ag bimetallic targets were acquired for four different compositions of 20-80, 30-70, 50-50, 80-20, respectively. A 600 µm optical fiber was used to couple the collected light to the spectrometer and ICCD, which was triggered by the delay generator output. Since the plasma emissions from metals as well as bimetals were strong the gate delay and gate width were kept long enough and were optimized for metals and bimetals separately. The acquisition parameters used in the experiments are summarized in table 1. Each spectrum was an accumulation (average) of 10 acquisitions taken from a fresh spot each time. Typically, 20 kinetic spectra were collected and were averaged at each time delay from the time resolved spectra and the final mean spectra (only the first five of the collected twenty spectra) were considered for the 2D correlation studies as shown in schematic of the figure 1(b). The averaging was done to minimize the matrix effect, if any. All the experiments were performed under similar conditions and the effect of ICCD (operating at $-25^0$ C) dark current was not considered in the analysis. The nonlinear variations in the intensity of the peaks with increasing the compositions are the primary motivation for this work to study the correlation between the transitions in the LIBS spectra. The coupling between the Au and Ag peaks for different compositions are illustrated at different ranges.

**2D Correlations Plots:**

A set of 'm' LIBS spectra of the system under systematic disturbance inducing change in the spectral intensities are represented by $y(\lambda_j, t_i)$. Where the discrete variable $\lambda_j$ for $j = 1,2,3..n$ related to the single spectrum represents the wavelength sampled over 'n' values.

The second variable $t_i$ for $i = 1,2,3…m$ represents the effect of applied perturbation on the system sampled over m values.

The dynamic spectrum $\tilde{y}(\lambda_j, t_i)$ in the time interval $t_1$ to $t_m$ is defined in the below equation as

$$\tilde{y}(\lambda_j, t_i) = y(\lambda_j, t_i) - \bar{y}(\lambda_j) \tag{1}$$

Where $\tilde{y}_j(v_j, t_i)$ is the averaged spectra used as reference in the calculation of the dynamic spectra. The reference spectra can be either first spectrum or last spectrum of the time resolved spectra and it can also be set to zero as well.[30] The reference spectrum is set to zero in our studies.

$$\bar{y}(\lambda_j) = \frac{1}{m}\sum_{i=1}^{m} y(\lambda_j, t_i) \tag{2}$$

The synchronous $\Phi(\lambda_1, \lambda_2)$ and asynchronous $\Psi(\lambda_1, \lambda_2)$ correlation spectra are written as

$$\Phi(\lambda_1, \lambda_2) = \frac{1}{m-1}\sum_{j=1}^{m} \tilde{y}_j(\lambda_1) \cdot \tilde{y}_j(\lambda_2) \tag{3}$$

Where $\tilde{y}_j$ is the spectral intensity at a point of variable $t_j$

$$\tilde{y}_j(\lambda_i) = \tilde{y}(\lambda_i, t_j) \qquad i = 1,2$$

$$\Psi(\lambda_1, \lambda_2) = \frac{1}{m-1} \sum_{j=0}^{m} \tilde{y}_j(\lambda_1) \cdot \sum_{j=0}^{m} N_{jk} \cdot \tilde{y}_k(\lambda_2) \tag{4}$$

Where $N_{jk}$ Hilbert-Noda transformation matrix

$$\begin{Bmatrix} 0 & if\ j = k \\ \frac{1}{\pi(k-j)} & otherwise \end{Bmatrix} \tag{5}$$

Spectra were measured with the perturbation in the system of interest like varying magnetic field, electrical field, thermal, chemical, optical, mechanical, or the system varying with time. The analysis divulges the similarities and dissimilarities in the spectra. Correlation analysis was applied on the time varying metal, bimetallic targets, and composition varying bimetallic targets LIBS spectra. The correlations with systematic variations would be easy to interpret after the two-dimensional (2D) spectroscopy. The resulting similarity spectra is called synchronous and dissimilarity spectra as asynchronous 2D spectra, which are complementary to each other. The asynchronous spectra in this case of study were discarded since they did not contain considerable variations. Figure 2 illustrates the time-resolved LIBS spectra of Aluminium [figure 2(a)], Copper, [figure 2(b)], and Brass [figure 2(c)] at five different times with regular interval of 1 µs and the same data was used for the 2D correlation studies of Aluminium [figure 2(d)], Copper, [figure 2(e)], and Brass [figure 2(f)], respectively.

**Results and Discussion**

Four different reference spectra were tried for the analysis as shown in figure 3(a) with the first spectrum (from the time series spectra). Figure 3(b) depicts the last spectrum (from the time series spectra) whereas figure 3(c) is the reference taken as zero (no spectrum) and figure 3(d) is the average spectrum (obtained from the time series spectra) used as reference. The variation in the amplitude and the width of the peaks are observed for different reference spectra. The peaks are broad and high when the first spectra in figure 3(a) was taken as reference and the width, height was reduced for both the cases of last spectrum as reference [figure 3(b)] and the zero as reference spectrum [figure 3(c)]. The intensities were too low when the reference spectrum used was an average of all the spectra [figure 3(d)]. This could be because of the flat baseline in LIBS spectra obtained from the gated spectrometer from which the emissions of plasma continuum cab be avoided. The amplitude of peaks in the synchronous correlation plot is a function of both the LIBS signal intensity of the correlated peaks and the strength of correlation from the changes between them. The contributions from the noise, which is random, is constant on to the diagonal of 2D synchronous plots because of lack of autocorrelation. Hence, 2D synchronous plots can be used to recover weak signal from a noisy background. As shown in figure 4, the linear spectra along the diagonal of matrix in the 2D analysis were considered to demonstrate the improvement in the signal. Figure 4(a) depicts the regular LIBS spectra of Aluminium in the spectral range of 303-312 nm and the corresponding 2D correlation spectra is shown in figure 4(b) while figure 4(c) illustrates the plot of diagonal of the 2D correlation analysis of the time resolved LIBS spectra.

**Correlation Studies**

The 2D correlation spectra of the LIBS have wavelength as both X-axis and Y-axis. The peaks on the diagonal represent the autocorrelation of the peak called auto-peaks and the off-diagonal peaks correspond to the correlation with the other peaks at the particular wavelength and are called cross-peaks. Python was used for the analysis and in the calculation of the synchronous and asynchronous spectra from the dynamic spectra with the average spectra taken as zero. The shape of the contour refers to the homogeneity of the peak broadening while circular contours indicate homogeneous broadening and elliptical contours indicate inhomogeneous broadening.

The noise in the 2D spectra was suppressed as the randomly varying noise does not have correlations with varying time but the signal with correlation is enhanced thus improving the signal to noise ratio in the 2D plots. Figure 2(d) shows that the atomic peaks 309.27 and 309.28 of Aluminium are correlated positively and the peak are tabulated in the table 2. In figure 2(e) the Copper peaks at 329.54 nm and 330.20 nm exhibited a good correlation and the identified peaks are tabulated table 3. The intensity of the cross-peaks shows how strongly the peaks are correlated. In figure 2(f) for Brass shows that the atomic peak of Copper at 324.75 nm is correlated to the atomic peak of Zinc at 334.5 nm this correlation was found to be stronger than the self-correlation of the Zinc peak. The LIBS peaks obtained for Brass are summarized in table 4. The correlation spectra of the Brass and Copper were similar in the range of 220-250 nm but were highly distinguishable in the range 320-340 as expected. The auto peak 327.39 nm was found to be weaker than the peak at 324.7nm and the cross peak suggesting a coupling between the transition at 327.39 nm and 324.7 nm.

The LIBS signal intensity of the atomic peaks was improved by 5 orders of magnitude in the case of Aluminium and the data is shown in figure 5 [5(a) representing Aluminium, 5(b) representing Copper, 5(c) representing Brass] while 4 orders of magnitude improvement was observed for bimetallic targets as shown in figure 5 data [(d) illustrating Au30Ag70, (e) illustrating Au50Ag50, and (f) illustrating Au80Ag20 targets data]. It is worth mentioning that the observed improvements in the SNR is for these particular sets of data and the experimental conditions mentioned earlier. We believe that similar improvements can be obtained for other LIBS data as well. Specifically, this analysis will be an excellent choice where the SNR is very poor. For example, standoff LIBS spectra of explosive molecules and that too in the single shot mode. However, further detailed studies are essential to confirm this hypothesis. Recently, Quaroni et al. reported the signal improved by 3 orders of magnitude for time resolved infrared spectra.[31] The correlated peaks were improved and the non-correlated noise was suppressed.

Furthermore, the widths of the peaks were reduced suggesting the improvement in resolution of the spectrum. This clearly demonstrates that multiple spectra under various conditions can be used to improve the quality of the spectra. Further studies are also necessary for exploring the application of this method to overcome the matrix effects in LIBS. The observed off-diagonal peaks were brighter than the diagonal peaks suggesting good correlations. The 2D correlations analysis from both the time resolved LIBS spectra of bimetal targets shown in figure 6(a) for Au30Ag70, figure 6(b) for Au50Ag50, and figure 6(c) for Au80Ag20. Five different gate delays with regular interval of 1 μs was used for the 2D correlation studies shown in the figure 6(d) for Au30Ag70, figure 6(e) for Au50Ag50, and figure 6(f) for Au80Ag20 targets, respectively. The target Au50Ag50 exhibited superior correlations than the other two. It is also observed that the first ionized peak of Ag at 241.32 nm does not show much variation with the change in the Ag percentage but the intensity of the off-diagonal peaks varies. The cross peak between the first ionized peak of Ag at 241.32 nm and the atomic peak of Au at 242.7 nm is stronger.

The varying composition of Ag/Au in Ag-Au alloy target was also used as a perturbing parameter and the obtained results are depicted in figure 7. The 2D analysis for three different compositions was compared between data obtained for 1 μs and 2 μs gate delays. The correlations for three different spectral regions (a) 240-250 nm, (b) 260-280 nm, and (c) 520-550 for the first μs and (d) 240-250 nm, (e) 260-280 nm, (f) 520-550 nm for the second μs after the pulse was incident, respectively, were compared. The correlations improved in the 240-250 nm range but they diminished in the other two ranges 260-280 nm and 520-550 nm. LIBS spectra of Au-Ag bimetal targets at three different compositions were used for the 2D correlation studies at two different gate delays, in the range 240-250 nm [figure 7(a)], 260-280 nm [figure 7(b)], 520-550 nm [figure 7(c)] at 1 μs and 240-250 nm [figure 7(d)], 260-280 nm [figure 7(e)], 520-550 nm [figure 7(f)], at 2 μs. The correlations between different compositions grew as the time increased in the range 240-250 nm and diminished in other two ranges. The identified LIBS peaks of both the Au and Ag are summarized in the table 5.

**Classification Studies**

Principal component analysis on the diagonal of the matrix in the 2D synchronous spectra and the regular LIBS spectra were performed and the results are depicted in figure 8. PCA maximizes the variance of the principal components[32] and thus dimensional reduction[33] is achieved and with fewer dimensional data we could be able to explain the system. The larger the variance, the greater is the amount of information the principal component contains. The cumulative explained variance provides the information of how many principal components should be included to describe the data accurately. The PCA score plot is shown in figure 8(a) and the variance plot is depicted in figure 8(b). The cumulative explained plot is illustrated in 8(c) for the regular LIBS spectra. The score plot [figure 8(d)], variance plot [figure 8(e)], and the cumulative variance plot [figure 8(f)] of the improved spectra were also compared. Here in this case removing the noise and improving the peaks intensity resulted in the improvement of the explained variance of each first three principal components and we could achieve 95% variance only with the first 3 components whereas in the case of regular spectra with noise it took more than 5 components to get 80%, which is much lower compared to earlier number.

The SNR of the spectra improved enormously in the case of the intense peaks in regular spectra and the highly correlated peaks with them. The correlation between the atomic peaks in Aluminium (303 – 310 nm range), Copper (320-340 nm range) and Brass (320-340 nm range) were visualized in figure 2. The coupling between the Zinc and Copper lines was strong in the case of Brass in the 320-340 nm range. The distinguishable coupling for different compositions of Au and Ag were visualized and the cross-peaks were observed to be strong for the Au50-Ag50 sample when compared to the remaining compositions in the range 240-250 nm. The Au-Ag correlations peaks were plotted from the time-resolved LIBS spectra for varied compositions and the observed cross-peaks were found to be strong in the 260-280 nm range, as depicted in the figure 6. These 2D plots could also help distinguish the samples of different compositions by filtering out the features which do not change. Even with the diagonal elements from the 2DCOS data we can improve the classification. Additional studies with different perturbation methods will be an enormously useful tool for a better understanding of the coupling between different transitions and also to improve the classification. This could be an interesting tool also to explore the molecular emissions in femtosecond laser-induced plasma[24,34,35] and in exploring the data for machine learning applications (such as PCA, artificial neural networks, convolution neural networks).

**Conclusions**

Primarily the SNR is enhanced in 2D spectroscopy since the randomly varying noise does not hold any correlations with time. The linear spectroscopic data cannot provide information regarding the correlations in the transitions. While the noise intensity is diminished the signal intensity was improved from the analysis. Additionally, 2DCOS appears to be a powerful tool in resolving the complex spectra as well. 2D correlation analysis simplifies the interpretation of correlations between the transitions in LIBS spectra and advantageous in visualising the time-dependent decay of different LIBS peaks and wavelength shift, if any. It is encouraging to report that our initial correlation analysis on the standoff LIBS data of polymers[18] has also provided enhancements in the SNR. Further detailed analysis is pending and will be reported elsewhere.

We summarize the important results from this work and suggestions for future works:

- These studies will provide a better understanding of the molecular peaks and their coupling, especially in the femtosecond LIBS spectra. For example, the molecular

- emissions such as CN, $C_2$, AlO, TiO in the LIBS spectra could add more insights to the LIBS analysis.

- We believe that 2DCOS studies combined with time-resolved LIBS spectra or other perturbation methods would result in better resolution with the same spectrometer, due to the spreading of data over a second dimension, improved resolution and SNR from which overlapped peaks can be distinguished, which are not possible otherwise.

- The same studies can be performed/repeated with other perturbations such as changing ICCD gain, gate delays, input laser energy, the distance at which the emissions are collected, compositions, etc. to get an improved understanding of the transitions and enhancements.

- In the case of the stand-off LIBS the SNR deteriorates as compared to nearfield case and this method can be used for improving it, especially in the case of explosives detection.[18,36,37]

**Conflicts of interest:** The authors declare no conflicts of interest.

**Acknowledgements:** Authors acknowledge DRDO, India for continuous financial support through Project #ERIP/ER/1501138/M/01/319/D (R&D) dated 27.02.2017. Authors also thank Dr. Chandu Byram for his assistance in some of the experiments.

**Table 1** Acquisition parameters used in the ns LIBS experiments.

| Samples | Gate delay | Gate width | Exposure time | ICCD Gain |
|---|---|---|---|---|
| **Aluminium, Copper, Brass** | 0.5 µs | 0.5 µs | 2 ms | 100 |
| **Au-Ag Bimetals** | 1 µs | 1 µs | 2 ms | 50 |

**Table 2** Identified LIBS peaks of Aluminium target

| Sl. No. | Wavelength Observed (nm) | Element | Ionization |
|---|---|---|---|
| 1 | 305.00 | Al | I |
| 2 | 305.71 | Al | I |
| 3 | 308.21 | Al | I |
| 4 | 309.27 | Al | I |
| 5 | 309.28 | Al | I |
| 6 | 358.65 | Al | II |
| 7 | 394.40 | Al | I |
| 8 | 396.15 | Al | I |
| 9 | 624.33 | Al | II |

**Table 3** Identified LIBS peaks of Copper target.

| Sl. No. | Wavelength Observed (nm) | Element | Ionization |
|---|---|---|---|
| 1 | 296.1.16 | Cu | I |
| 2 | 324.7.54 | Cu | I |
| 3 | 327.3..96 | Cu | I |
| 4 | 329.0.54 | Cu | I |
| 5 | 330.7.95 | Cu | I |
| 6 | 333.7.84 | Cu | I |
| 7 | 465.1.12 | Cu | I |
| 8 | 510.5.54 | Cu | I |
| 9 | 515.3.24 | Cu | I |
| 10 | 521.8.20 | Cu | I |
| 11 | 529.2.52 | Cu | I |
| 12 | 578.2.13 | Cu | I |

**Table 4 Identified LIBS peaks of Brass target.**

| Sl. No. | Wavelength observed (nm) | Element | Ionisation |
|---|---|---|---|
| 1 | 324.7.54 | Cu | I |
| 2 | 327.3.96 | Cu | I |
| 3 | 329.054 | Cu | I |
| 4 | 330.7.95 | Cu | I |
| 5 | 333.7.84 | Cu | I |
| 6 | 330.2.58 | Zn | I |
| 7 | 330.2.93 | Zn | I |
| 8 | 334.5.01 | Zn | I |
| 9 | 334.5.56 | Zn | I |
| 10 | 334.5.93 | Zn | I |
| 11 | 468.0.13 | Zn | I |
| 12 | 472.2.15 | Zn | I |
| 13 | 481.0.53 | Zn | I |

**Table 5 Identified LIBS peaks of Au-Ag bimetallic target**

| Sl. No. | Wavelength Observed (nm) | Element | Ionisation |
|---|---|---|---|
| 1 | 241.32 | Ag | II |
| 2 | 242.7 | Au | I |
| 3 | 243.77 | Ag | II |
| 4 | 247.39 | Ag | II |
| 5 | 267.5 | Au | I |
| 6 | 276.7 | Au | I |
| 7 | 328.13 | Ag | I |
| 8 | 338.32 | Ag | I |
| 9 | 520 | Ag | I |
| 10 | 546 | Ag | I |
| 11 | 768 | Ag | I |

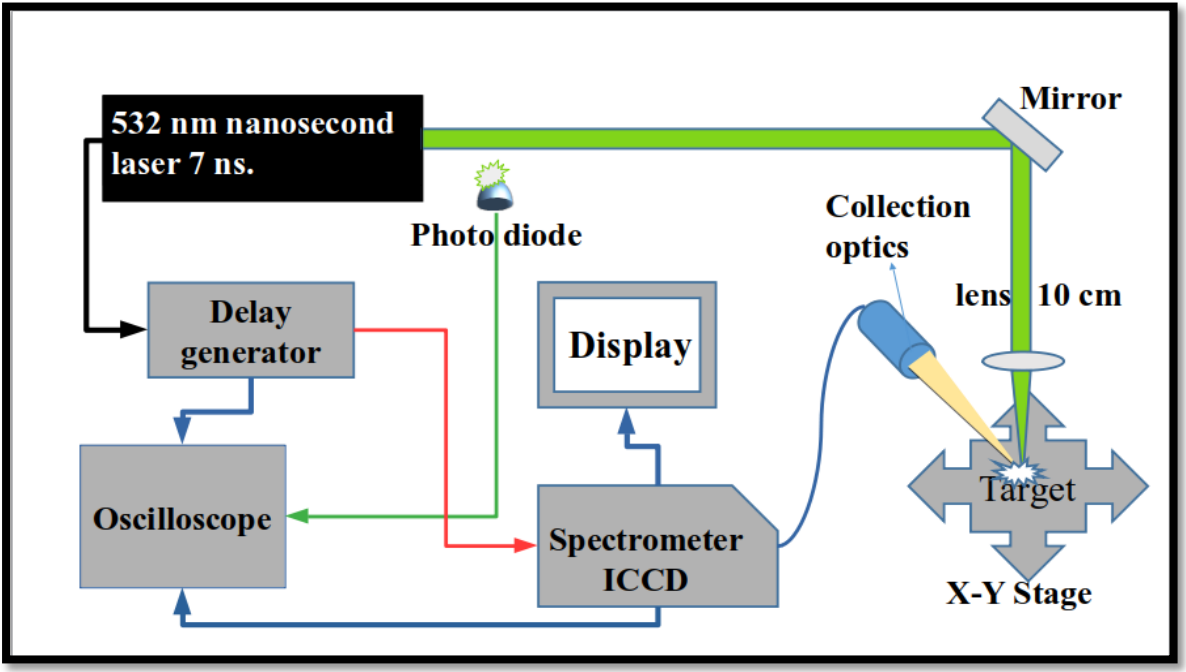

**(a)**

**Fig. 1(a)** Schematic of the nanosecond LIBS experimental setup used for metals, alloys, and bimetallic targets.

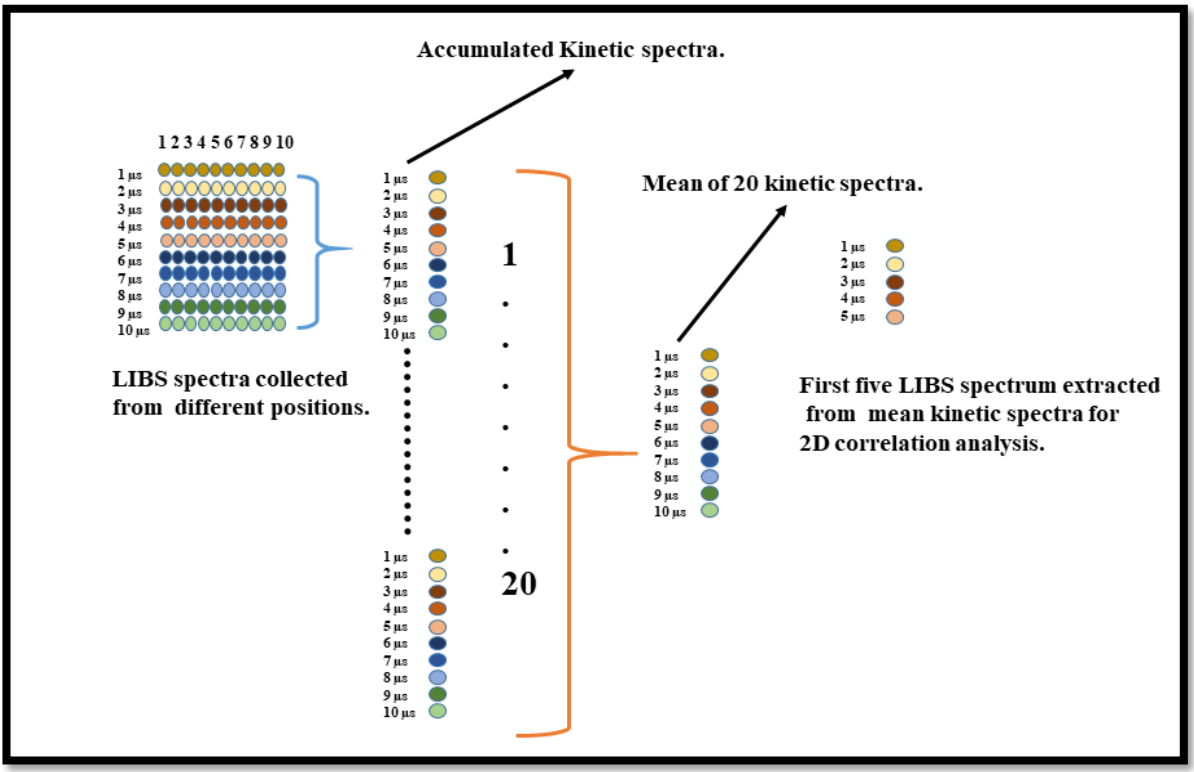

**(b)**

**Fig. 1(b)** The steps followed in data acquisition and data preparation for the analysis.

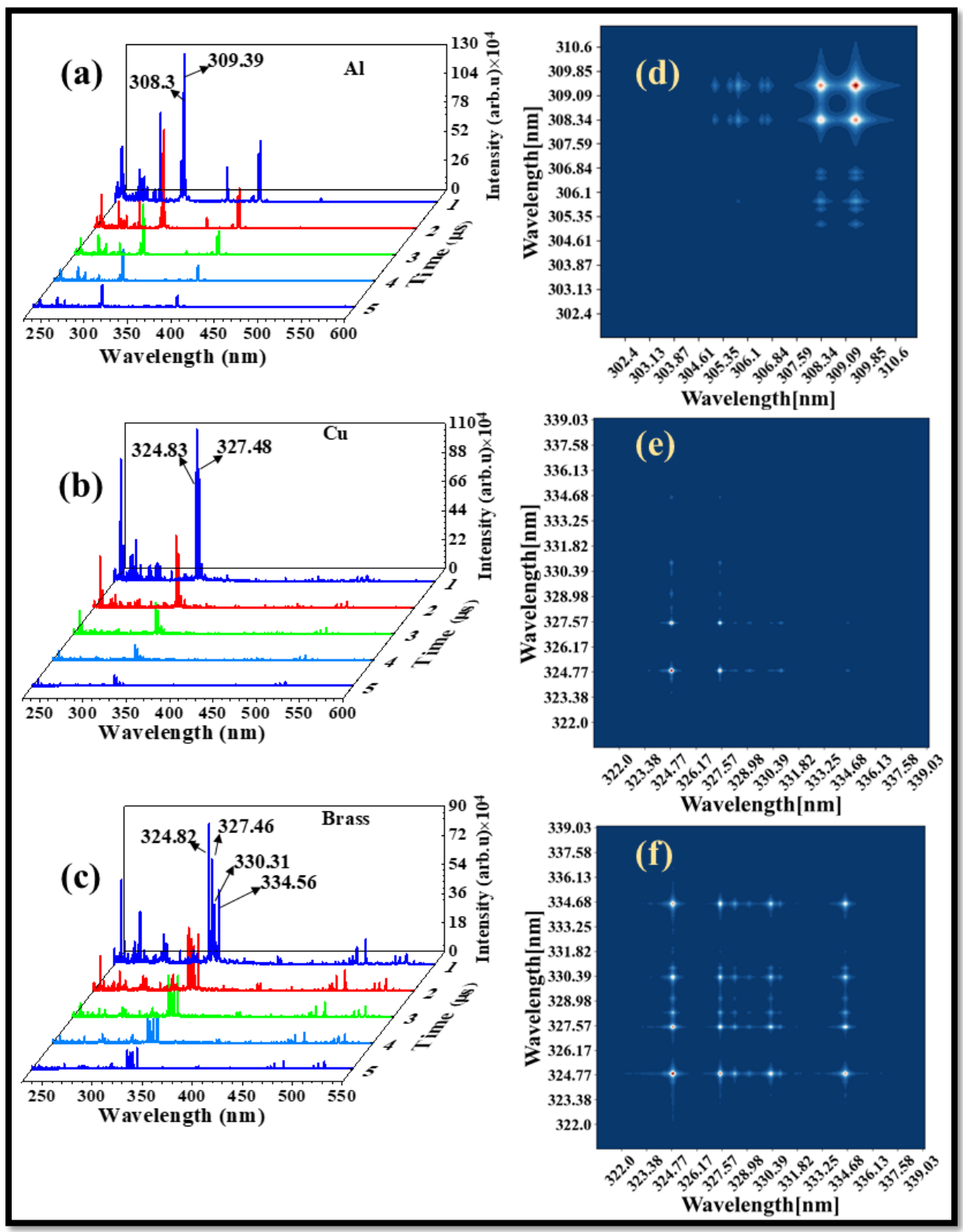

**Fig. 2** The time-resolved LIBS spectra of (a) Aluminium (b) Copper (c) Brass at five different gate delays with regular interval of 1 μs. These were used for the 2D correlation studies for (d) Aluminium (e) Copper (f) Brass targets, respectively.

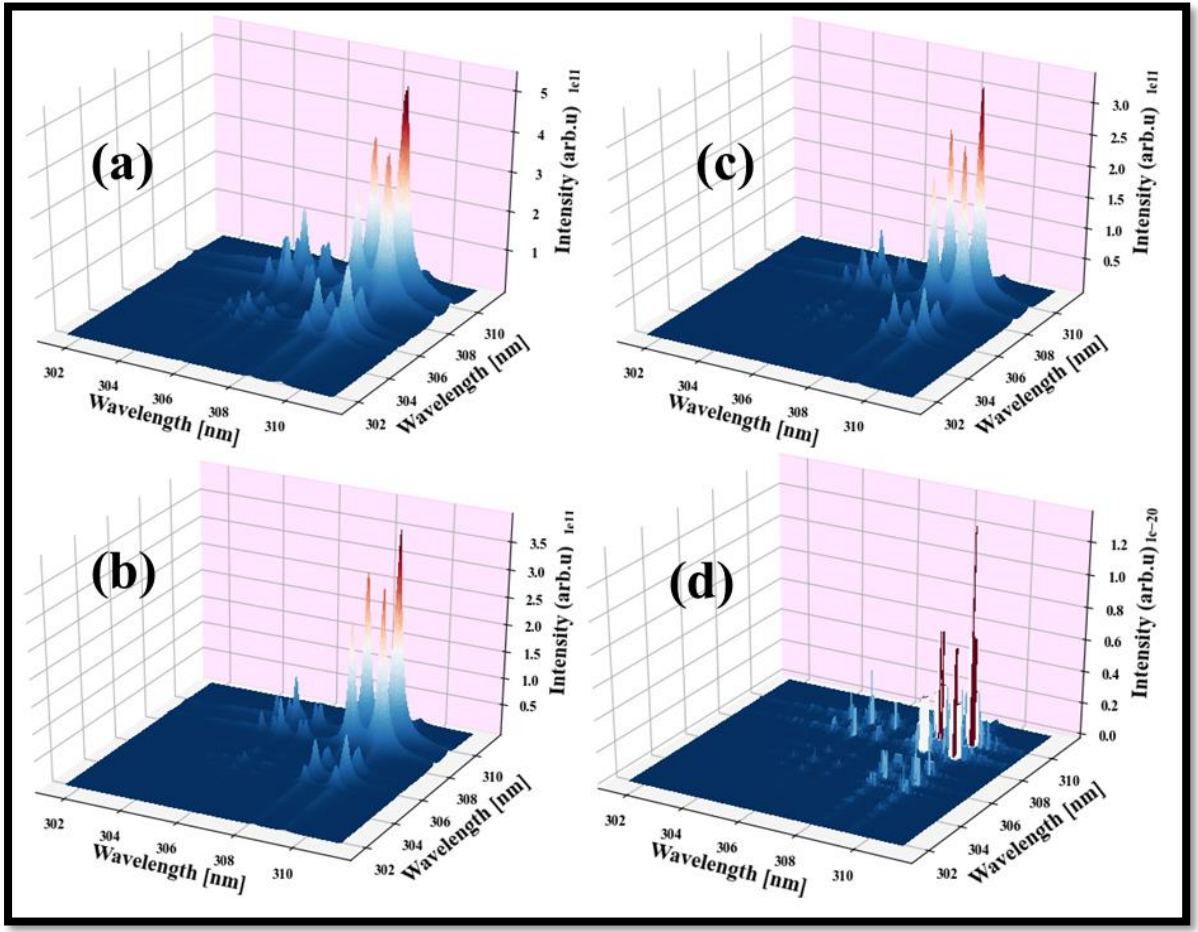

**Fig. 3** 2D correlation analysis performed with four different reference spectra (a) with the first collected spectrum of the time-resolved spectra (b) last collected spectrum of the time-resolved spectra (c) with reference taken as zero and (d) average spectrum of the time-resolved spectra as reference.

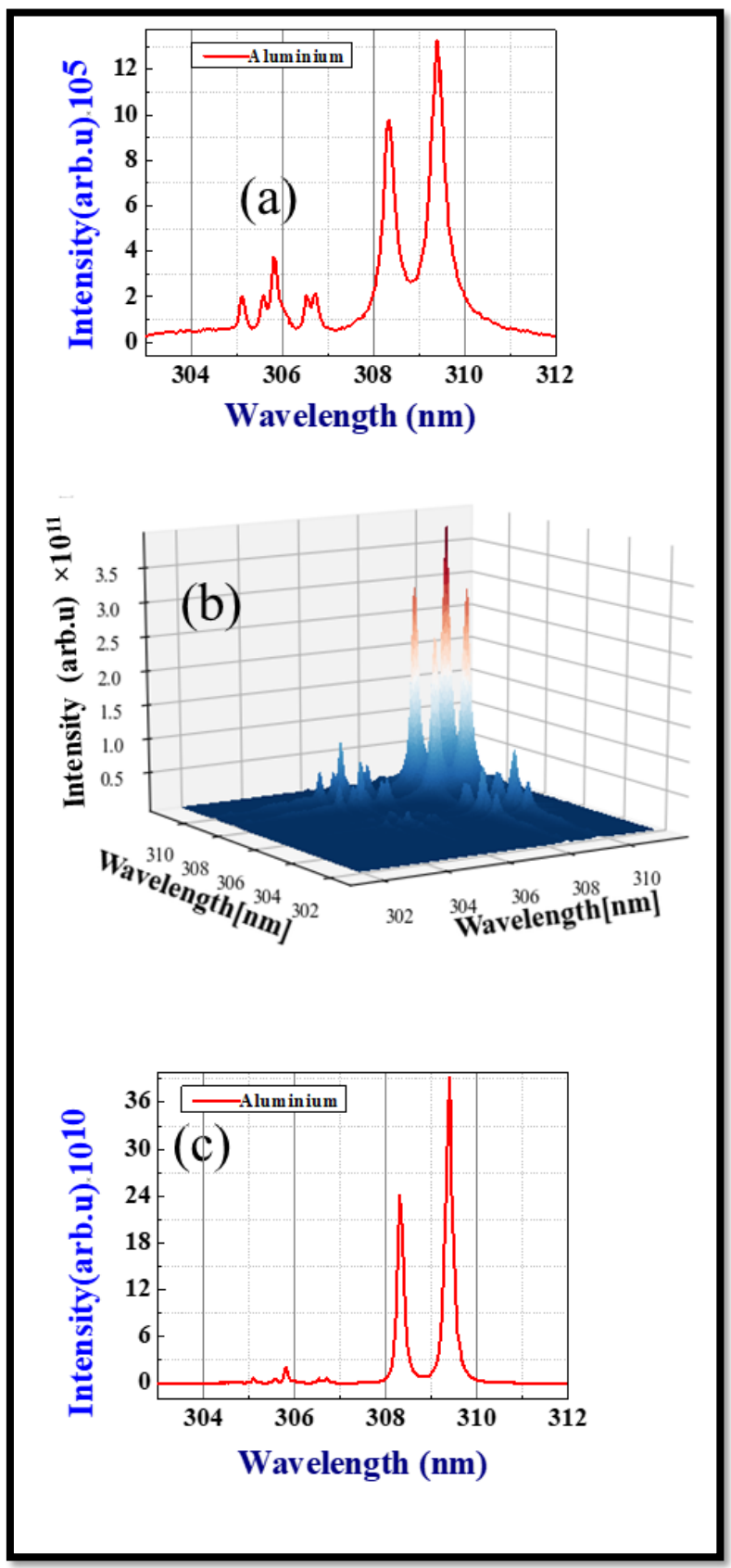

**Fig. 4** The regular LIBS spectra of (a) Aluminium target in the spectral range of 303-312 nm and the corresponding (b) 2D correlation spectra and (c) diagonal of the 2D correlation analysis of the LIBS spectra is plotted.

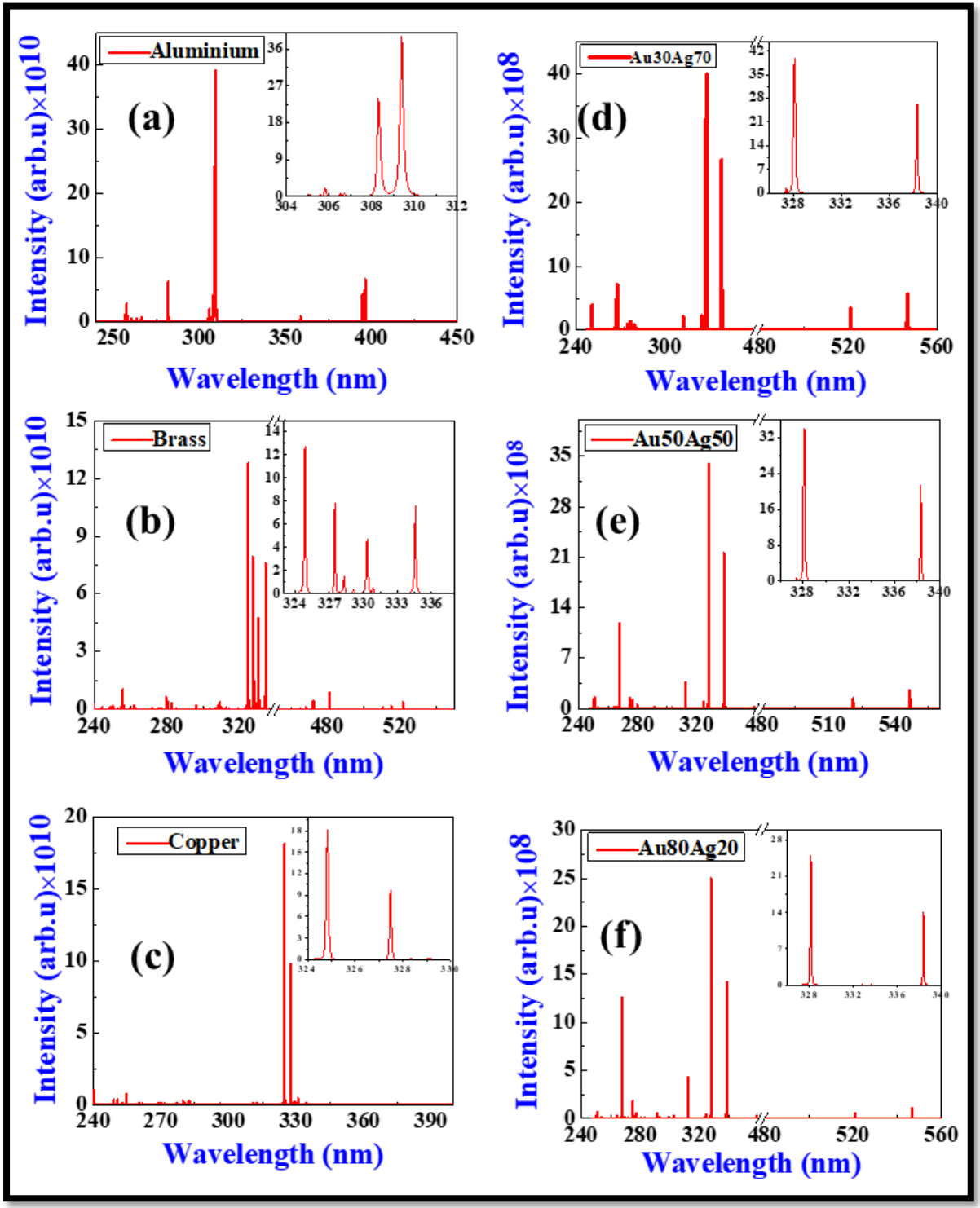

**Fig. 5** The diagonal of the 2D correlation analysis on LIBS spectra is plotted. The signal to noise ratio is improved enormously for all the samples (a) Aluminium (b) Copper (c) Brass (d) Au30Ag70, (e) Au50Ag50 (f) Au80Ag20 targets, respectively.

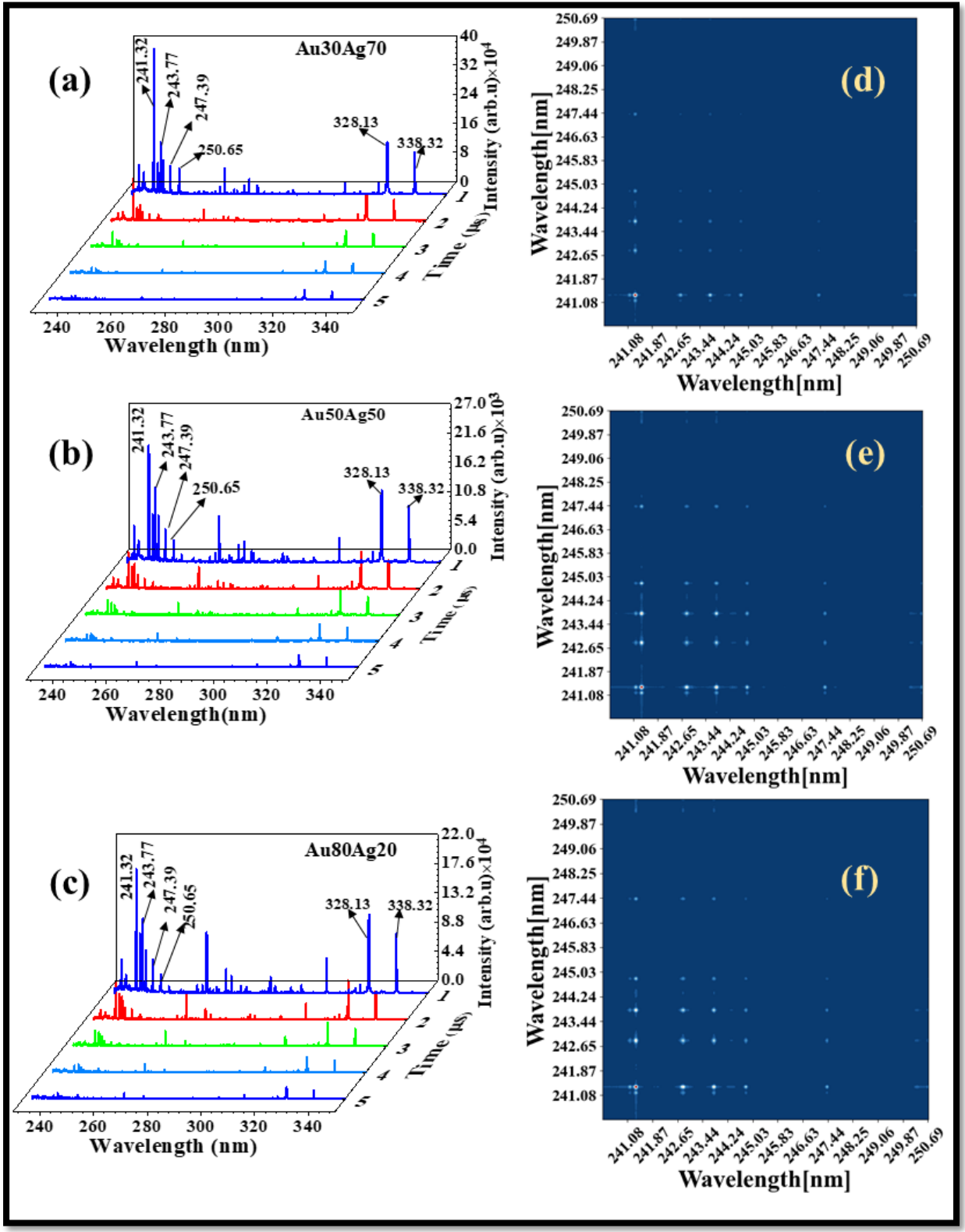

**Fig. 6** Time-resolved LIBS spectra of (a) Au30Ag70, (b) Au50Ag50, (c) Au80Ag20 bimetallic targets at five different gate delays with regular interval of 1microsecond is used for the 2D correlation studies on the time-resolved spectra of (d) Au30Ag70, (e) Au50Ag50, (f) Au80Ag20 targets, respectively.

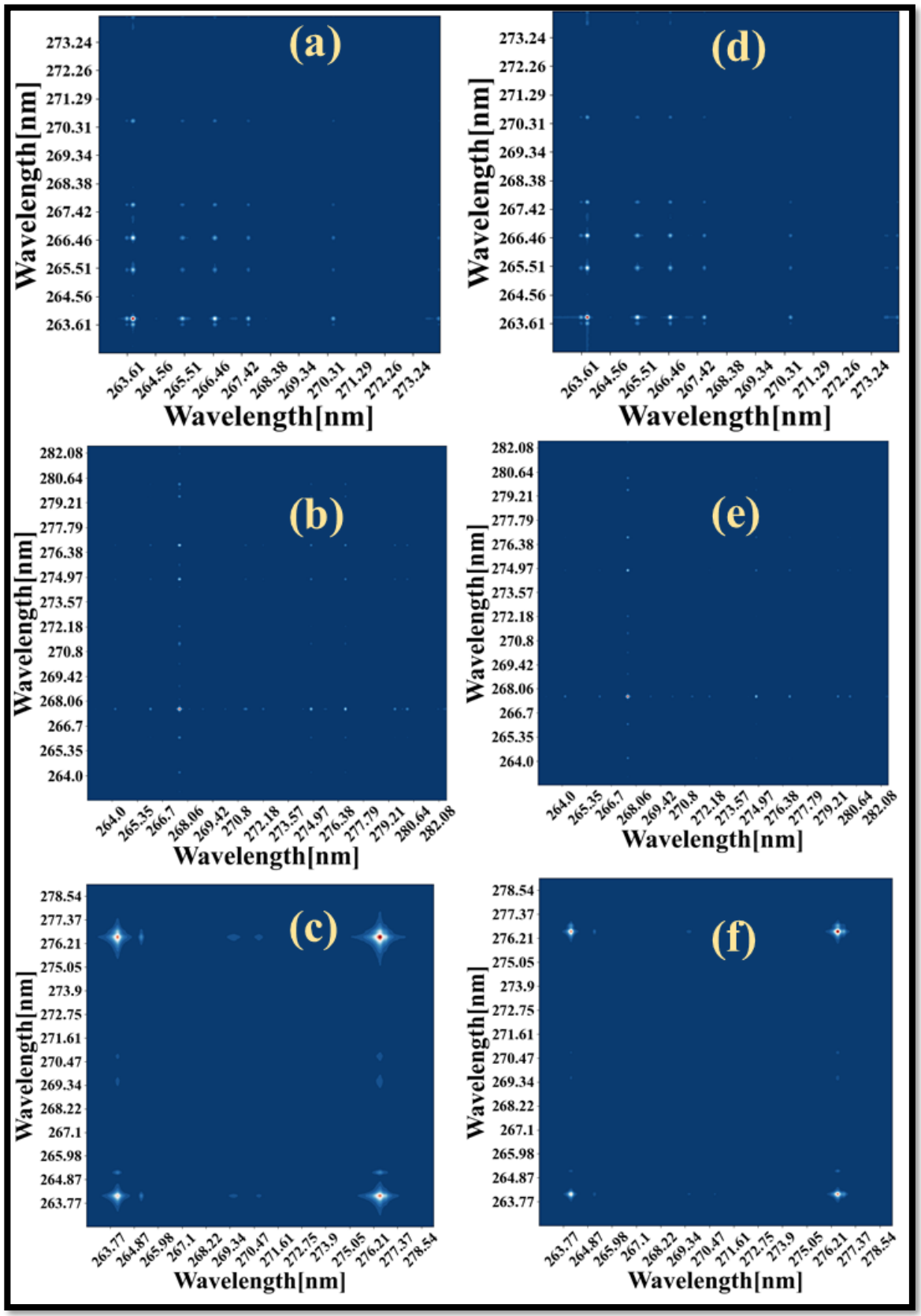

**Fig. 7** LIBS spectra of Au-Ag bimetal at three different compositions were used for the 2D correlation studies at two different gate delays in the range (a) 240-250 nm (b) 260-280 nm (c) 520-550 nm at 1 μs and (d) 240-250 nm (e) 260-280 nm (f) 520-550 nm at 2 μs.

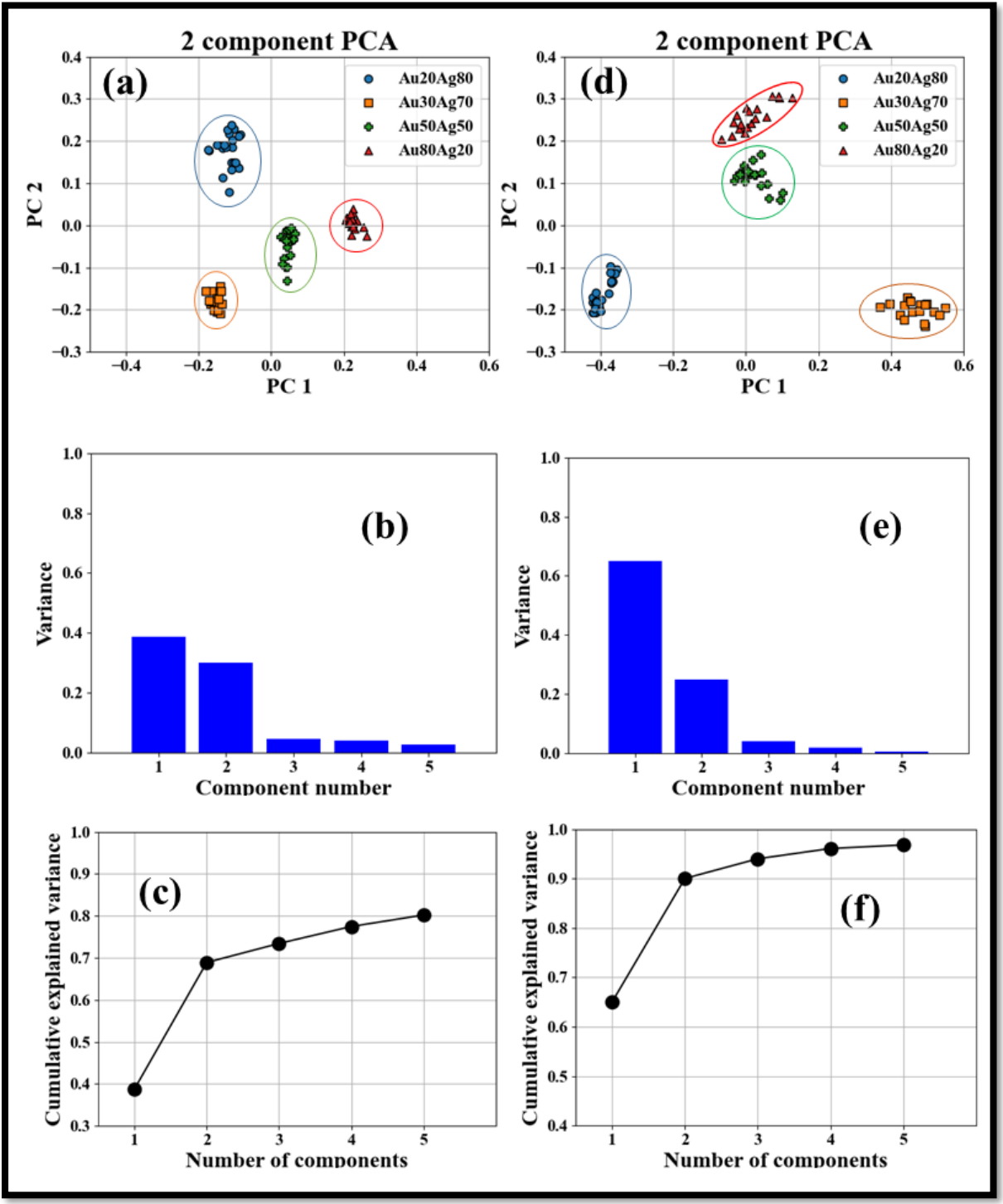

Fig. 8 PCA on the regular spectra of bimetal targets and the improved diagonal of the 2D correlation analysis (a) PCA score plot (b) variance plot (c) cumulative explained variance on regular spectra and (d) PCA score plot (e) variance plot (f) cumulative explained variance on the signal to noise improved spectra. Solid line is a guide to the eye.